\documentclass[aps,prd,twocolumn,letterpaper,showpacs,floatfix]{revtex4}
\pdfoutput=1

\usepackage[squaren]{SIunits}
\usepackage{amssymb,amsmath}
\usepackage{graphicx}
\usepackage{dcolumn}
\newcolumntype{.}{D{.}{.}{-1}}
\usepackage[breaklinks=true]{hyperref}

\begin{document}

\graphicspath{{plots/}}

\newcommand{\Eref}[1]{Eq.~(\ref{#1})}
\newcommand{\Fig}[1]{Fig.~\ref{#1}}
\newcommand{\Figure}[1]{Figure~\ref{#1}}
\newcommand{\Sec}[1]{Sec.~\ref{#1}}
\newcommand{\Tab}[1]{Tab.~\ref{#1}}

\newcommand{\Pizza}{\textsc{pizza\ }}
\newcommand{\Coconut}{\textsc{coconut\ }}

\newcommand{\D}{\partial}
\newcommand{\Dc}{\nabla}
\newcommand{\Rmd}{\rho}
\newcommand{\Sed}{\epsilon}
\newcommand{\Sent}{h}
\newcommand{\Crmd}{D}
\newcommand{\Ced}{\tau}
\newcommand{\Cmom}{S}
\newcommand{\Sam}{l_\varphi}

\newcommand{\Pcmom}{\delta \Cmom}
\newcommand{\Pcrmd}{\delta \Crmd}
\newcommand{\Psam}{\delta \Sam}
\newcommand{\Ps}{\delta s}
\newcommand{\Pw}{\delta w}
\newcommand{\Pv}{\delta v}
\newcommand{\Px}{\delta x}
\newcommand{\Psed}{\delta \Sed}
\newcommand{\Prmd}{\delta \Rmd}

\newcommand{\Sg}{\sqrt{g}}
\newcommand{\Vc}{\sqrt{\gamma}}
\newcommand{\Lf}{W}
\newcommand{\Lapse}{\alpha}
\newcommand{\A}{A}
\newcommand{\B}{b}
\newcommand{\C}{C}
\newcommand{\En}{\nu}
\newcommand{\Enc}{\En_c}
\newcommand{\Ndir}{\psi}

\newcommand{\Half}{\frac{1}{2}}
\newcommand{\Inv}[1]{\frac{1}{#1}}
\newcommand{\Flt}[2]{\ensuremath{{#1}\times 10^{#2}}}

\title{Inertial modes of rigidly rotating neutron stars in Cowling approximation}
\author{Wolfgang Kastaun}
\affiliation{Department of Theoretical Astrophysics, Universit\"at
T\"ubingen, 72076 T\"ubingen, Germany}
\date{\today}

\begin{abstract}
In this article, we investigate inertial modes of rigidly rotating
neutron stars, i.e.\ modes for which the Coriolis force is dominant.
This is done using the assumption of a fixed spacetime
(Cowling approximation).
We present frequencies and eigenfunctions for a sequence of stars
with a polytropic equation of state, covering a broad range of rotation rates.
The modes were obtained with a nonlinear general relativistic hydrodynamic
evolution code.
We further show that the eigenequations for the oscillation modes
can be written in a particularly simple form for the case of arbitrary fast,
but rigid rotation.
Using these equations, we investigate some general characteristics of inertial
modes, which are then compared to the numerically obtained eigenfunctions.
In particular, we derive a rough analytical estimate for the frequency
as a function of the number of nodes of the eigenfunction,
and find that a similar empirical relation matches the numerical results
with unexpected accuracy.
We investigate the slow rotation limit of the eigenequations, obtaining
two different sets of equations describing pressure and inertial modes.
For the numerical computations we only considered axisymmetric modes,
while the analytic part also covers nonaxisymmetric modes.
The eigenfunctions suggest that
the classification of inertial modes by the quantum numbers of
the leading term of a spherical harmonic decomposition
is artificial in the sense that the largest term is not strongly
dominant, even in the slow rotation limit.
The reason for the different structure of pressure and inertial modes
is that the Coriolis force remains important in the slow rotation limit
only for inertial modes.
Accordingly, the scalar eigenequation we obtain in that limit
is spherically symmetric for pressure modes, but not for inertial modes.
\end{abstract}

\maketitle
\section{Introduction}
Neutron star oscillations have become an interesting field of research
because they are possible sources of gravitational waves,
which might be directly detected in the near future.
Also, some are believed to undergo rotational instabilities where
energy is transferred from the rotation of the star to the oscillation
and gravitational radiation, which also carries away angular momentum.
Such mechanism would not only be a strong source of gravitational waves,
but also a possible explanation for the observed limitation of the rotation rate
of neutron stars.
Should gravitational waves ever be detected, we would gain some insight
into the behavior of cold, dense matter, see \cite{Kokkotas98}.
For this, one needs accurate theoretical models.

There are different families of oscillations:
Spacetime modes (w-modes) are oscillations of the spacetime itself with 
weak coupling to the matter, see \cite{Schutz92}, 
and only exist in the framework of general relativity.
For pressure modes (f- and p-modes),
the restoring forces are mainly pressure and gravitation.
They exist in all stars, regardless of entropy gradient and rotation rate.
If the star has a radial specific entropy gradient,
there also exist g-modes, where the restoring
force is buoyancy.
For inertial modes, the Coriolis force,
gravitation, and pressure are equally important restoring forces.
We will only consider the case of isentropic stellar models, 
for which no g-modes exist; 
according to \cite{Lockitch2000} inertial modes
of non-nonbarotropic stars behave qualitatively different.
Formally, inertial modes only exist in rotating stars,
but for arbitrary low rotation rates.
In the nonrotating limit, their frequency goes to zero and they become
stationary currents.

So far, astrophysical interest has been focused on a 
subclass of nonaxisymmetric inertial modes called $r$-modes,
because in the absence of viscosity and nonlinear mode coupling 
effects, they can undergo rotational-gravitational instabilities 
at any rotation rate, see \cite{Andersson98}.
Recent Newtonian studies \cite{Lin2006,Teukolsky2004,Arras2003} argue
that the instabilities are in fact strongly suppressed by mode coupling.
However, for the axisymmetric modes we extracted numerically,
no rotational instabilities exist.

Our knowledge on neutron star oscillations in general relativity
is mainly based on perturbative studies.
In \cite{Thorne67I,Thorne69II,Thorne69III,Thorne69IV,Thorne70V,Thorne73VI,Detweiler83,Detweiler85}
oscillations of nonrotating stars are investigated in full relativity.
Pulsations of slowly rotating stars are studied in
\cite{Lockitch2000,Kokkotas02,Kokkotas03,Kojima92,Kojima93,Lockitch03}.
In \cite{Yoshida97,Yoshida05}, 
nonaxisymmetric pressure modes and $r$-modes are studied for
arbitrary rotation rates, but using the Cowling approximation

Most results are limited to the slow rotation approximation,
because in that limit it is feasible to decompose the perturbed
quantities into spherical harmonics, reducing the equations
to one spatial dimension.
Linearized evolution equations in full relativity
using this formalism can be found
in \cite{Kokkotas02,Kokkotas05,Kojima92}.
In \cite{Kokkotas03}, inertial modes are investigated using the
additional assumption of a fixed spacetime.

As shown in the above references, rotation couples
spherical harmonic contributions of different quantum number $l$.
This yields an infinite system of equations,
which is truncated to $l<l_m$ to obtain numerical solutions.
For the case of inertial modes however, a problem arises:
as detailed in \cite{Kokkotas03},
the resulting equations admit solutions only outside certain frequency
bands, which change their location and number with the chosen
truncation $l_m$ in an apparently non-converging fashion.

In the analytic part of our work, we try to investigate the inertial mode problem
from a different angle and directly study the
two dimensional perturbed eigenequations.
Additionally, we do not restrict ourselves to the slow rotation limit from
the beginning, although it is used for some of our conclusions.
We do however rely on the assumption of rigid rotation.

Oscillations of neutron stars have also been studied using
nonlinear relativistic hydrocodes, which are not restricted to the slow rotation
approximation.
In \cite{SAF2004,Dim2001}, pressure modes frequencies are computed
within the Cowling approximation, and in \cite{DSF2006} using the
conformal flatness approximation. The latter also contains some inertial
mode frequencies, but without identification of the eigenfunction.
In \cite{Font2001}, a simulation of one inertial mode with
a high amplitude in the nonlinear regime has been carried out in full relativity.
To our knowledge, there are no inertial mode eigenfunctions
of rapidly rotating stars available in the literature so far.

This article is organized as follows:
In \Sec{sec_ana} we present the eigenequations and investigate the
general characteristics of inertial modes,
particularly in the slow rotation limit, which is derived
in \Sec{sec_slow}.
In \Sec{sec_num}, we present numerically extracted eigenfunctions
and frequencies,
which are then compared to the analytic expectations.
\section{Analytic properties}
\label{sec_ana}
In this section, we present the eigenequations in Cowling approximation
for the case of rigid rotation.
Although there is no known analytic solution, we deduce some general properties
of inertial modes.
They will be used in \Sec{sec_num} to check and explain the numerical results.

We use the following notation:
$\Rmd$ is the rest mass density in the fluid restframe,
$\Sed$ the specific internal energy, excluding restmass,
$\Sent = 1 + \Sed + P / \Rmd$ the specific relativistic enthalpy,
$c_s$ the speed of sound,
$u^\mu$ the 4-velocity of the fluid, 
$v^i$ the 3-velocity measured by an observer with worldline
normal to the hypersurface of constant coordinate time,
$\Lf$ the corresponding Lorentz factor,
and $w^i = u^i / u^0$ the advection speed of the fluid with respect to 
the coordinates. 
The 3-metric is denoted by $g_{ij}$, its volume element
by $\Vc = \sqrt{\det(g_{ij})}$, and the Lapse function by $\Lapse$.
We further use the conserved mass density $D=\Vc\Lf\Rmd$,
the conserved momentum density $\Cmom_i = \Vc \Lf^2 \Rmd \Sent v_i$,
and the specific angular momentum $\Sam = \Cmom_\varphi / \Crmd$.
The angular velocity of the star measured by a distant observer is denoted
by $\Omega$.

We use a cylindrical type coordinate system $(t,d,z,\varphi)$ in which
the stationarity of the spacetime and the axisymmetry are manifest,
i.e.\ all background quantities are invariant under translation in
$t$- and $\varphi$-direction.
We further require the coordinate system to be corotating,
i.e.\ $w^i=0$. This is always possible for rigid rotation,
but otherwise not.
Spacelike indices are denoted by letters $i,j \in \{d,z,\varphi\}$,
while indices $l,k \in \{d,z\}$ denote components in the meridional plane.
In perturbation equations, the perturbed quantities are generally
prefixed with a $\delta$,
while everything else refers to the unperturbed background model.
If not noted otherwise, geometric units where $G=c=1$ are used.
\subsection{Linearized equations}
\label{sec_eeq}
To obtain perturbed evolution equations, we linearize the hydrodynamic evolution
equations on arbitrary curved spacetimes in the formulation given in \cite{Kastaun06},
which read
\begin{eqnarray}
  \D_t \Crmd &=&
    - \D_j \left( \Crmd w^j \right) ,\\
  \D_t \Cmom_i &=&
    -  \D_j \left[\Cmom_i w^j
    +\Lapse\Vc\left( P-\hat{P} \right) \delta^j_i \right]
  \quad\bigm\vert_{Q}.
\end{eqnarray}
The second equation is only valid at the arbitrarily chosen
point $Q$, for which the function $\hat{P}$ is defined by
\begin{equation}
  \hat{P}(Q) = P(Q),\qquad
  \frac{\Lapse}{\hat{\Lf}} h(\hat{P}) = \text{const},
\end{equation}
where $\hat{\Lf}$ is the Lorentz factor that belongs to
a constant advection speed $\hat{w}^i \equiv w^i(Q)$.
In this formulation,
the geometrical source terms which are usually written in terms
of derivatives of the metric tensor,
are combined in the derivative of $\hat{P}$.
For the following, it is not necessary to understand this
formulation, which is explained in detail in \cite{Kastaun06}.
The energy equation also derived there is redundant
due to the assumption of adiabatic evolution.

The above equations are derived assuming the stress energy tensor
of an ideal fluid and the conservation of restmass.
Here we further assume an isentropic stellar model and adiabatic evolution,
both satisfying the same one-parametric equation of state (EOS).
For the unperturbed stellar model, 
we generally assume axisymmetry and rigid rotation.
As shown in \cite{Kastaun06}, any stationary, isentropic,
and rigidly rotating stellar model has to satisfy
\begin{equation}\label{eq_kappa}
  \kappa \equiv \frac{\Lapse}{\Lf} h = \text{const}.
\end{equation}
Using this as well as the axisymmetry of the unperturbed background model,
a straightforward, but lengthy calculation leads to the evolution
equations for the perturbation
\begin{eqnarray}
  \D_t \Pcrmd &=&
    -\D_i \left( \Crmd \Pw^i \right),
  \label{eq_evol_mass}\\
  \D_t \Pcmom_\varphi &=&
    - \D_i \left( \Cmom_\varphi \Pw^i \right)
    - \kappa \Crmd \D_\varphi \Ps,
  \label{eq_evol_cam}\\
  \D_t \Pcmom_l &=&
    \Crmd \left( - \kappa \D_l \Ps + \Pw^\varphi \D_l \Sam \right),
  \label{eq_evol_mom}
\end{eqnarray}
where $\Ps \equiv \delta\ln(\Sent)$.
Note the above equations are valid only in corotating coordinates.
The system is closed by the algebraic relations
\begin{eqnarray}
  \Pcmom_l &=&
    \Crmd \Sent \frac{\Lf}{\Lapse} \Pw_l,
  \\
  \Pcmom_\varphi &=&
     \Cmom_\varphi \left(
      \left(1+\Inv{c_s^2} \right)  \Ps
      +  \left(1 + v^2 \right) \frac{\Lf^2}{\Lapse v_\varphi} \Pw_\varphi
    \right),
  \\
  \Pcrmd &=&
    \Crmd \left(
      \Inv{c_s^2} \Ps + \frac{\Lf^2}{\Lapse} v_\varphi \Pw^\varphi
    \right).
\end{eqnarray}
\Eref{eq_evol_mass} describes the conservation of rest mass.
Because of the axisymmetry of the background model,
there is a conserved angular momentum $\Cmom_\varphi$,
described by \Eref{eq_evol_cam}.
To understand \Eref{eq_evol_mom}, we compute its Newtonian limit,
obtaining
\begin{eqnarray}
  \D_0 \delta v^d &=&
    - \D_d \Ps + 2\Omega \delta v^\varphi d,
  \label{eq_newt_accel_d}\\
  \D_0 \delta v^z &=&
    - \D_z \Ps.
\end{eqnarray}
The term $2\Omega \delta v^\varphi d$, which originated
from the term $\Pw^\varphi \D_l \Sam$ in \Eref{eq_evol_mom}, is nothing
but the Coriolis force component acting in $d$-direction,
caused by the change of velocity in $\varphi$-direction.
The forces due to the perturbed balance of pressure
and gravitational force are all contained in the
term $-\D_l \Ps$.
Note the latter is only possible when using \Eref{eq_kappa}.

We now assume harmonic time dependence for all perturbations $\delta Z$.
Because of the axisymmetric background, we also assume harmonic dependence on
$\varphi$, writing
\begin{equation}\label{eq_harm_ans}
  \delta Z(d,z,\varphi,t) = \delta Z(d,z) e^{i (\omega t +m \varphi)}.
\end{equation}
Note $\omega$ is the frequency with respect to coordinate time
in the corotating frame; the frequency in a nonrotating frame
is shifted by $m\Omega$.
After some straightforward computations, we obtain the eigensystem
\begin{eqnarray}
 0 &=&
  \Psam
  + \Px^l \D_l \Sam
  + m \frac{\kappa}{\omega} \Ps,
  \label{eq_ee_sam}\\
 0 &=&
  \Pcrmd + \D_l \left( \Crmd \Px^l \right)
  + m \Crmd \frac{1}{\omega} \Pw^\varphi,
  \label{eq_ee_mass}\\
 0 &=&
  \Px^k \left(\omega \frac{\Lf}{\Lapse} \right)^2 \A_{kl}
  - \D_l \Ps
  - 2 q \Omega |v| \frac{\Lf}{\Lapse} \B_l \Ps.
  \label{eq_ee_mom}
\end{eqnarray}
Here,
$\Px^k = \Pw^k / (i\omega)$ is the fluid displacement vector,
and
\begin{eqnarray}
  \A_{kl} &=&
    g_{kl} - \Inv{\En^2} \B_k \B_l,
  \\
  \B_l &=&
    \frac{\Lapse}{\Lf} \frac{|v|}{2\Omega} \D_l \ln \Sam,
  \\
  \En &=&
    \frac{\omega}{2\Omega},
  \\
  q &=&
    1 + m \frac{\kappa}{\omega \Sam}.
\end{eqnarray}
The vector $b^l$ depends weakly on the rotation rate;
in the Newtonian limit, we obtain $b^l=(1,0)$.
The above system of equations has real coefficients.
The physical (real-valued) solutions for
the quantities $\Ps$, $\Pw^\varphi$, and $\Px^k$ therefore
oscillate in phase, which is spatially constant.
The velocity components $\Pw^k$ in the meridional plane
are phase-shifted by $\pi / 2$ with respect to $\Ps$.

If $\En^2 \neq \B^l\B_l$, $\A^k_l$ can be inverted and \Eref{eq_ee_mom} can be
solved for $\Px^l$.
Together with \Eref{eq_ee_mass}, we obtain a second order scalar master
equation for $\Ps$.
\begin{equation}\label{eq_master}
  0 =
    \omega^2 \Crmd \eta \Ps +
    \D_l \left( \frac{\Lapse^2}{\Lf^2} \Crmd \C^{lk} \D_k \Ps \right),
\end{equation}
where
\begin{eqnarray}
  \C^{kl} &=&
    g^{kl} - \frac{\mu^2}{\En^2} \B^k \B^l, 
  \\
  \Enc^2 &=&
    \B^l\B_l, \qquad
    \mu^2 = \frac{\En^2}{\Enc^2 - \En^2}, 
  \\
  \eta &=&
    \mu^2 v^2 q^2 + \Inv{c_s^2}
    - \Inv{\Crmd} \D_l
      \left( \Crmd \frac{|v| \Lapse \mu^2}{2\Omega \Lf \En^2} q \B^l \right).
\end{eqnarray}
A very similar formulation, valid also for differential rotation,
is given in \cite{Lindblom92}.
The principal part of the equation is determined by the
tensor $\C^k_l$,
which has the eigenvalues $\{1,-\mu^2\}$.
One can therefore expect qualitatively different families of solutions
for the cases $\En>\Enc$ and $\En<\Enc$.
From the Newtonian limit of \Eref{eq_ee_sam} and \Eref{eq_ee_mom}, it
is easy to show that for axisymmetric modes, 
the ratio between the  $d$-components of the Coriolis force
and the total net force acting on a fluid element is given by $1/\En^2$.
Identifying the relativistic terms that reduce to the Coriolis force in the
Newtonian limit, we obtain in the relativistic case that
for $\En<\Enc$, the Coriolis force is not only the dominant
restoring force, it even has to be counterbalanced by the pressure force.
In $d$-direction, pressure forces and fluid acceleration then become
{\em antiproportional}.
If $\En \gg \Enc$ on the other hand, the Coriolis force becomes unimportant.
This is exactly the criterion that distinguishes inertial modes and pressure
modes.
\subsection{Slow rotation limit} \label{sec_slow}
In the following, we investigate the slow rotation
limit $\Omega \to 0$ of the eigenequations.
It is important to note that for inertial modes,
the oscillation frequency $\omega$ is proportional
to the rotation rate $\Omega$.
Therefore, just ignoring terms of higher order in $\Omega$
in the eigenequations would be wrong.
First, we have to replace $\omega$ by $2\Omega\En$.
Then we take the limit $\Omega \to 0, \En \to \En_0$,
where $\En_0$ is some finite value.

For pressure modes on the other hand, the oscillation
frequency $\omega$ approaches a finite value $\omega_0$ for
$\Omega \to 0$.
Hence the proper slow rotation limit is given by
$\Omega \to 0, \omega \to \omega_0 > 0, \En \to \infty$.

In this case, $\C^{kl} \to g^{kl}$.
\Eref{eq_master} in the slow rotation limit then becomes
\begin{equation}
0 =
  \left(  \frac{\omega_0^2}{c_s^2} -m^2 \frac{\Lapse^2}{g_{\varphi\varphi}} \right) \Ps
  + \frac{1}{\Crmd} \D_l \left( \Lapse^2 \Crmd g^{lk} \D_k \Ps \right).
\end{equation}
Using \Eref{eq_harm_ans} and the axisymmetry of the background, 
this can also be written in the covariant form
\begin{equation}
  0 =
    \omega^2_0\frac{\Rmd}{c^2_s} \Ps
    + \left(\nabla_i \left(\Lapse^2 \Rmd\right)\right)
      \nabla^i \Ps
    + \Lapse^2 \Rmd \Delta\Ps.
\end{equation}
Neglecting rotational corrections of second order in the rotation
rate, the quantities $\Rmd$, $c_s$, and $\Lapse$ are given by the spherically 
symmetric TOV solution, see \cite{OV39,Hartle67}.
The above equation becomes invariant under rotation
and can be solved by the ansatz
\begin{equation}
  \Ps(r,\theta,\varphi) =
    \Ps(r) Y_l^m(\theta,\varphi).
\end{equation}
It has been shown that there is an infinite
discrete set of solutions for given $l$, $|m| \leq l$,
exactly one for a given number $n$ of radial nodes.
The frequency grows without bound with increasing $n$ or $l$.

For the inertial modes, such an ansatz is not possible.
The reason is that in the slow rotation limit for inertial
modes, we obtain
\begin{equation}
\C^{kl} \to \C^{kl}_0 =
    g^{kl} - \Inv{\Enc^2 - \En_0^2} \B^k \B^l.
\end{equation}
The second term is finite,
which means that other than the star profile,
the equations themselves are not even approximately
spherically symmetric.
This is reflecting the fact that the Coriolis force,
which is a directional force,
remains dominant for arbitrary slow rotation rates.

In perturbative mode calculations like \cite{Kokkotas03},
the scalar quantities are
usually decomposed into spherical harmonic contributions
$\sum f_n^l(r) Y_l^m(\theta)$.
For pressure modes, this is a natural choice since in the nonrotating
limit, one such term is enough to describe the solution,
while for moderate rotation rates, the others remain small
corrections.

For inertial modes on the other hand,
one cannot a priori expect that the above decomposition
is more natural than any other expansion,
e.g.\ as twodimensional Fourier series.
As will be shown in \Sec{sec_num}, the angular dependency
of axisymmetric inertial mode eigenfunctions
does indeed not bear resemblance to spherical harmonics.
This is not to say that spherical harmonic decomposition
is a bad choice, but to stress that inertial modes
have a completely different structure than pressure modes.

The eigenequations we obtain for inertial modes in the slow rotation
limit are given by
\begin{eqnarray}
  0 &=&
    \bar{\eta} m^2 \Ps
    + \Inv{\Crmd} \D_l \left(\Crmd \Lapse^2 \C^{kl}_0 \D_k \Ps \right),
  \\
  \bar{\eta} &=&
    \frac{\mu_0^2 \Lapse^2}{g_{\varphi\varphi}}
    - \Inv{\Crmd} \D_l \left(
        \Crmd \frac{\Lapse^2}{\sqrt{g_{\varphi\varphi}}}
        \frac{\mu_0^2}{\En_0} \B^l
      \right).
\end{eqnarray}
In particular for $m=0$, the equation is quite simple and might
be suitable for direct numerical solution.
Note however that the unknown $\En$ enters the equations in a
nonlinear manner as a downside.

From \Eref{eq_ee_sam} to \Eref{eq_ee_mom}, it follows
that in the inertial mode slow rotation limit,
\begin{equation}\label{eq_ordering}
\mathcal{O}(\Ps) = \mathcal{O}(\Omega \Pw^l)
=\mathcal{O}(\Omega \Pw^\varphi) = \mathcal{O}(\Omega^2 \Px^i).
\end{equation}
The density perturbation thus vanishes for a finite velocity
amplitude in the slow rotation limit.
The physical interpretation is that pressure and Coriolis
force are proportional for inertial modes, and the latter is
proportional to the rotation rate. 
The mass current has to become divergence-free.
For axisymmetric oscillations, our numerical results show that 
this is realized  by fluid motions which correspond to 
a sum of convection-like motions in the meridional plane
and differential rotation,
both of course with a harmonic time dependence 
and phase shifted by $\pi/2$.

Note that for an arbitrary small but fixed velocity amplitude, 
the fluid displacement vector
$\Px^i$ goes to infinity in the slow rotation limit.
However this does not mean the linear approximation is not valid anymore
to describe such an oscillation, since the Eulerian perturbations
remain all finite.
Only for the description of the movement of a fluid element
one would have to integrate the velocity field along its path,
which will result in the aforementioned motions.
For slow rotation rates, one can thus have arbitrary many turnarounds
during one oscillation period.

We now briefly discuss what happens in the presence of 
a small, but finite entropy gradient.
Lets assume the fluid moves adiabatically with exactly the same velocity field as 
in the isentropic case.
The density perturbation would then also be the same.
The Eulerian perturbation of specific entropy would be 
proportional to the fluid displacement vector, if the latter is small, but remain
finite in any case, since the fluid elements only move around inside the star as argued before.
Due to the entropy perturbation, there is an additional pressure perturbation
compared to the isentropic case.
The size of this perturbation has to be compared to the pressure perturbation
corresponding to the one of the isentropic inertial mode oscillation.
If it is much smaller, it can be regarded as a small correction to the isentropic inertial
mode oscillation, which is still an approximate solution. 
If it is much bigger on the other hand, the oscillation must have a completely different
structure than in the isentropic case.

If the entropy gradient is small enough, 
we thus expect three different regimes for finite amplitude inertial mode oscillations:
a small amplitude regime where the structure of the modes changes drastically,
a medium amplitude regime where the modes are similar to the isentropic case,
and the nonlinear regime.
The first case is captured by linear perturbation theory, where all oscillations
are arbitrary small by definition. 
Indeed, \cite{Lockitch2000} found qualitatively different solutions in the case of an entropy
gradient.
Wether there is a medium regime before the nonlinear regime starts
depends on the size of the entropy gradient 
as well as on the rotation rate; demanding that the velocity 
perturbation is bound by some finite value, the maximum pressure perturbation 
of isentropic inertial modes is proportional to the rotation rate.
Thus we expect that finite amplitude inertial mode oscillations 
of rapidly rotating stars are more robust to entropy gradients than in slowly
rotating stars.
\subsection{Boundary conditions}
\label{sec_bc}
We now investigate the behavior of the eigenequations at the surface.
For this, we assume that $\Ps$ and its first and second derivatives
remain finite at the surface.
This assumption is compatible with our numerical results.
It is easy to show that if there is a surface at all,
$\left|\D^l \Rmd\right| \gg \Rmd$ near the surface.
Neglecting the terms directly proportional to $\Rmd$ in
\Eref{eq_master}, we arrive at
\begin{eqnarray}
  \omega^2 \Ps V &=&
    \C^{kl}\left(\D_l \Rmd\right) \left( \D_k \Ps \right),
  \\
  V &=&
    \frac{\Lf}{\Lapse} \left(
      \frac{|v| \mu^2}{2\Omega \En^2 } q \B^l \D_l \Rmd
      - \frac{\Rmd}{\Lapse \Vc c_s^2}
    \right).
\end{eqnarray}
For the following, we assume that $\Rmd/c_s^2$ remains finite
near the surface, as it is the case for the polytropic EOS used in our numerical
computations.
For the axisymmetric case where $q=1$, the slow rotation limit for
inertial modes then leads to the condition
\begin{equation} \label{eq_ndir}
  \Ndir^k  \D_k \Ps  = 0, \qquad
  \Ndir^k \equiv  \C^{kl} \D_l \Rmd
\end{equation}
at the surface.

To investigate the behavior of $\Ndir^l$, we decompose
$\D_l\Rmd$ into the normalized eigenvectors of $\C^l_k$,
which are given by
$e_\top^l = b^l / |b|$, with eigenvalue $-\mu^2$,
and $e_\perp^l$ defined by
$e_\perp^l b_l = 0$, $|e_\perp|=1$,
corresponding to the eigenvalue $+1$.
The eigenvectors $e_\top$ and $e_\perp$ are roughly
(in the Newtonian limit exactly) orthogonal,
respectively parallel, to the rotation axis.
For small $\En$ and hence small $\mu$, $\Ndir^l$
will point in the direction of $e_\top$ away from the axis,
except near the equatorial plane, where $e_\top^l \D_l\Rmd \approx 0$.
In that case, nodes crossing the surface at some
distance from the equatorial plane will
be approximately parallel to the rotation axis.
\subsection{Small wavelength limit}
\label{sec_wave}
Since the numerically extracted inertial modes shown in \Sec{sec_num}
expose increasingly regular patterns with growing number of nodes,
it is instructive to investigate the limit of small
wavelengths.
Let's assume there is a solution for which the first and second
derivatives of $\Ps$, normalized to its maximum amplitude,
are much greater than the normalized derivatives of
the background quantities.
Treating the background quantities as constant in the neighborhood
of some point inside the star,
\Eref{eq_master} becomes
\begin{eqnarray}
  0 &=&
    U \Ps
    + \C^{kl} \D_k\D_l \Ps,
  \\
  U &=&
    \omega^2  \frac{\Lf}{\Lapse^2}\eta.
\end{eqnarray}
This can be solved by the ansatz
\begin{equation}\label{eq_wave}
  \Ps =
    \Ps_0 \: e^{i k^l x_l}.
\end{equation}
We now decompose $k^l$ into normalized eigenvectors of $\C^l_k$,
see \Sec{sec_bc},
writing $k^l = k_\top e_\top^l + k_\perp e_\perp^l$.
It follows a dispersion relation
\begin{equation}
  \mu^2 =
    \frac{k_\perp^2 - U}{k_\top^2}.
\end{equation}
Globally, the eigenfunctions cannot be described as plane waves.
Nevertheless, we can roughly estimate the local second derivatives
in each direction from the number of nodes along the
(entire) rotation axis and equatorial plane,
setting $k_\perp \approx n_r \pi / (2 R_p)$ and
$k_\top \approx n_e \pi / R_e$, where $R_p$ and $R_e$ are the polar
and equatorial coordinate radius of the star.

For the axisymmetric case,
we can also neglect $U$
in the slow rotation approximation for inertial modes,
which yields the relation
\begin{equation}\label{eq_approx}
  \mu \approx \frac{R_e}{2R_p} \frac{n_r}{n_e}
  ,\qquad
  \En =  \frac{\Enc}{\sqrt{1+\left(1/\mu\right)^2}}.
\end{equation}
We stress that the above approximation is quantitatively very bad.
For example, $\En$ is a global constant while $\Enc$ varies
by 30~\% in the stellar models we investigated.

However,
qualitatively we expect that frequencies of inertial modes
increase with the number
of nodes along the rotation axis,
but {\em decrease} with the number of nodes
along the equatorial plane.
We further expect that for high order modes $\En$ is bounded
by some $\bar{\Enc}$, which is of similar magnitude
as the values of $\Enc$ inside the star.
Finally, for small $\mu$, there should be nodes roughly parallel
to the rotation axis.
\subsection{Beyond Cowling}
In the following we will briefly discuss the emission of gravitational
waves and the quality of the Cowling approximation,
for inertial modes in slowly rotating stars.

A consequence of \Eref{eq_ordering} for estimating the gravitational
waves emitted by inertial modes is that the dominant contribution
in the quadrupole formula is given
by the current terms, not the density terms.
This can be seen as follows:
the second time derivative of the mass quadrupole is given by
$
  \ddot{I} =
    \left< \ddot{\Rmd} \right>_m \sim
    \Omega^2 \left<\Prmd\right>_m
$,
where the brackets denote the integral operators given in \cite{Thorne80},
while for the current quadrupole we have
$
  \ddot{S} =
    \left< \D_t^2(\Rmd v^i) \right>_c =
    \Omega^2 \left< v^i \Prmd + \Rmd \Pv^i \right>_c
$.
Since $\Prmd \sim \Ps$, $v\sim \Omega$ and
$\mathcal{O}(\Ps) = \mathcal{O}(\Omega \Pw^i)$,
we get
$\mathcal{O}(\ddot{I}) \sim \mathcal{O}(\Omega^2 \Ps)$, but
$\mathcal{O}(\ddot{S}) \sim \mathcal{O}(\Omega \Ps)$.

However, in the slow rotation limit the gravitational waves
produced by inertial modes become negligible
since their frequency goes to
zero; for the luminosity, we obtain
$L \sim |\dddot{S}|^2 \sim \Omega^6 \left<\Pv^i\right>_c^2$,
or $\mathcal{O}(L) \sim \mathcal{O}(\Omega^6 E_k)$,
where $E_k$ is the kinetic energy in the oscillation.
Therefore gravitational radiation should not play a role
for the dynamics of inertial mode oscillations of slowly 
rotating stars.

Assuming that dissipative and mode-coupling effects
limit the velocity perturbation amplitude to $|\Pv^i|< v_M$,
it follows an upper bound $L_m \sim v_M^2 \Omega^6$ for the
gravitational wave luminosity produced by any inertial mode.

Besides gravitational radiation, the Cowling approximation
neglects already on the Newtonian level the perturbation of 
the gravitational potential.
It is unclear how this affects inertial modes.
Although the density perturbation and hence the perturbation
of the potential goes to zero in the slow
rotation limit for a fixed velocity amplitude, 
the pressure perturbation and the Coriolis force do the same 
and cannot be neglected.
For pressure modes, it is known that the Cowling approximation
can be quite inaccurate, as shown in \cite{DSF2006}.
In \cite{Finn88}, it has been argued that
for the case of pressure modes of nonrotating stars
the Cowling approximation becomes exact
in the low frequency limit.
Note however that the Cowling approximation discussed there
is a refinement where the metric component $g_{rt}$ is also
perturbed;
see \cite{Lindblom90} for a comparison of the two variants.
Anyway, these results cannot be directly carried over to inertial modes
of rotating stars.
We thus believe that the Cowling approximation \textit{might} be more 
accurate for inertial modes than for pressure modes,
but this can only be quantified by a fully relativistic study.
\section{Numeric results} \label{sec_num}
The aim of this section is to give a qualitative overview how inertial
modes actually look like, to investigate the influence of the rotation rate,
and to provide accurate frequencies for reference in future works.
\subsection{Method}
To obtain frequencies and eigenfunctions, we perturb an
equilibrium stellar model with a small trial perturbation and
then evolve the general nonlinear hydrodynamic equations in time,
using the \Pizza code described in \cite{Kastaun06,thesis}.
We use the Cowling approximation, i.e.\ the spacetime is kept
fixed.
This saves a lot of resources, since we only need to
evolve the interior of the star,
but not a huge volume of surrounding spacetime.
The amplitudes of the perturbations are generally in the linear regime,
with velocities in the range $10^{-7}$ to $10^{-4}\usk c$.
We use the nonlinear \Pizza code instead of a linear
code only because it is available and well tested.

\begin{figure}
  \includegraphics[]{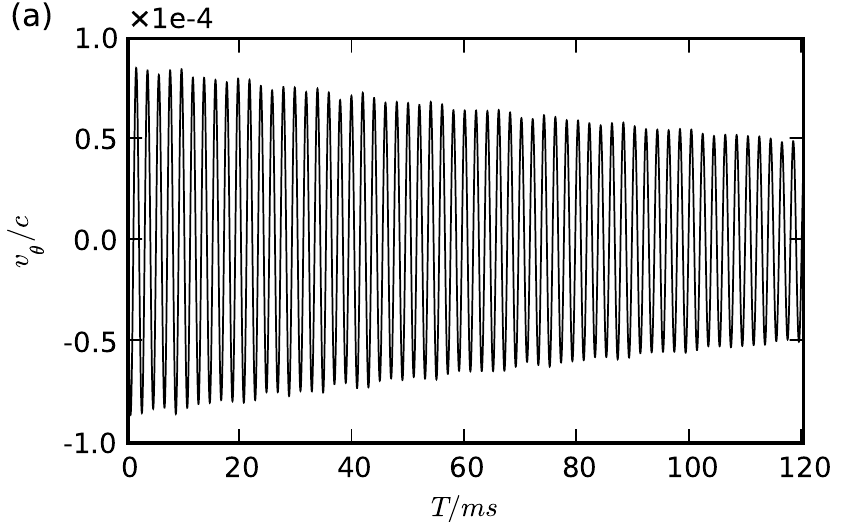}
  \includegraphics[]{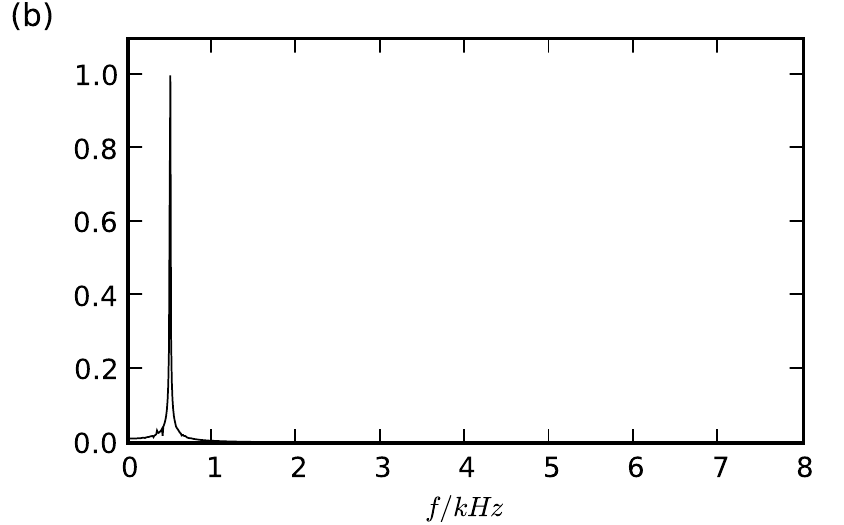}
  \caption{\label{fig_series}
    Time evolution (a) and Fourier spectrum (b) of velocity in
    $\theta$-direction at $r=3\usk\kilo\meter$,
    $\theta=\pi/2$, for a simulation of stellar model BU3, which was
    perturbed using the eigenfunction of inertial mode $i_{22}$.
    The resolution of the simulation is $150\times 150$ points.
    The damping is purely numerical.
  }
\end{figure}

From the Fourier spectra of the density and velocity variations
at some sample point inside the star, we determine the frequencies
of the excited oscillations.
We then perform a Fourier analysis for one of those frequencies
at every point inside the star, to obtain a first approximation
to the eigenfunctions.
The real part of the complex result is then used as initial perturbation
in a second simulation.
This step is repeated until other modes are sufficiently suppressed,
like in the time evolution and spectra shown in \Fig{fig_series}.
This method called mode recycling has already been used in \cite{DSF2006}.

As shown in \Sec{sec_eeq}, the velocity components
in the meridional plane are phase shifted by $\pi /2$ against the perturbations
of specific energy and velocity in $\varphi$-direction.
Apart from the global complex phase, the eigenfunctions are purely real.
To obtain the eigenfunctions
from the complex-valued result of the Fourier analysis, we first
compute a weighted mean phase by
\begin{equation}
  \phi_0 =
      \frac{\int \Vc \Rmd (\arg(\Psed_c) \mod \pi) \usk d^2x}
      {\int \Vc \Rmd  \usk d^2x },
\end{equation}
where $\Psed_c$ is the result of the Fourier analysis for the
specific energy perturbation.
The eigenfunctions are then obtained from $\Psed = \Re(e^{-i\phi_0} \Psed_c)$,
$\Pw^\varphi = \Re(e^{-i\phi_0} \Pw^\varphi_c)$, and
$\Pw^l = \Im(e^{-i\phi_0} \Pw^l_c)$.

\begin{figure}
  \includegraphics[]{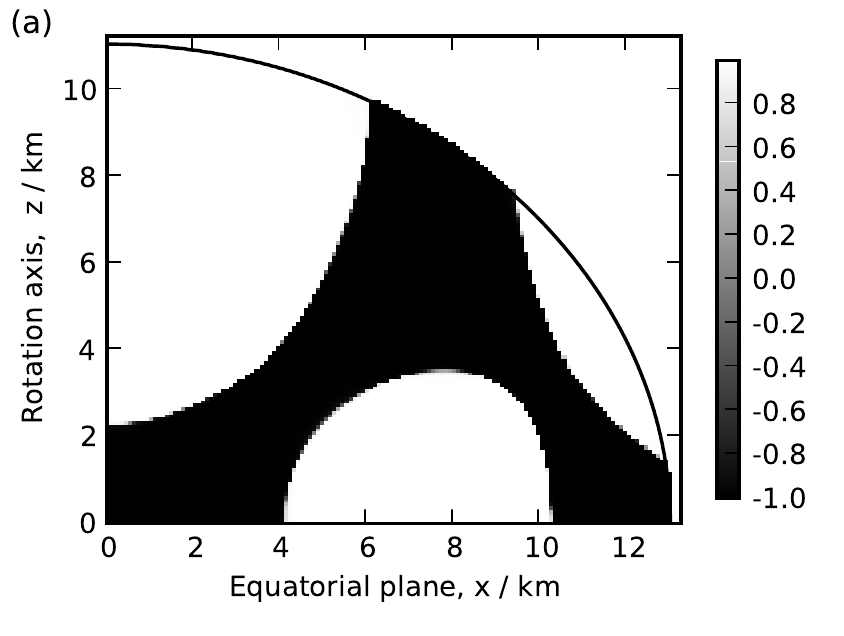}\\
  \includegraphics[]{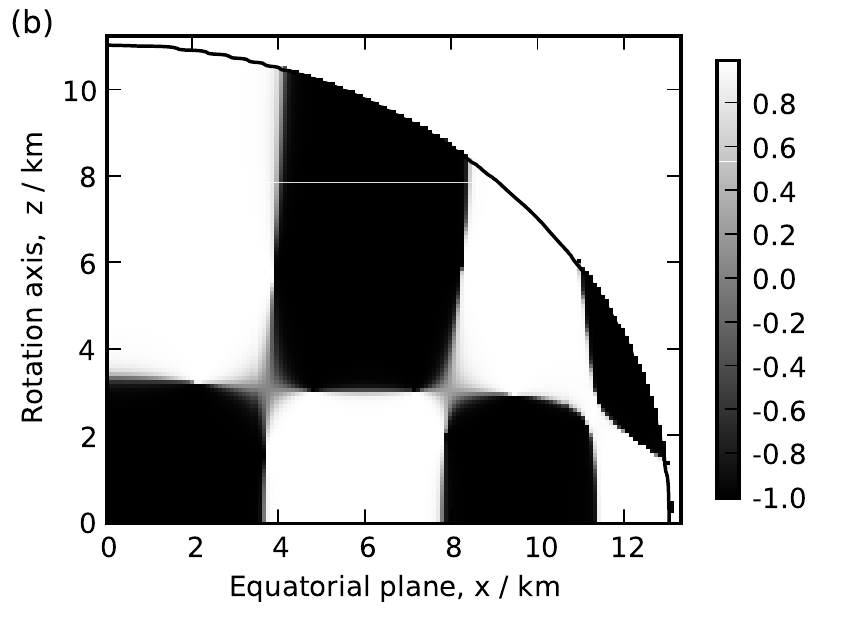}
  \caption{\label{fig_phase}
    Uniformity of the global complex phase under ideal circumstances (a)
    and for a problematic case (b).
    Color coded is the quantity $\cos(\arg(\Psed_c) - \phi_0)$.
  }
\end{figure}

We also compute the quantity $\cos(\arg(\Psed_c) - \phi_0)$, which should
result in $\pm 1$, with a sign flip at the nodes of the eigenfunction $\Psed$.
This is used to estimate the accuracy of the extracted mode, in
particular to determine if more mode recycling steps are necessary.
Examples how this measure looks like in our computations are shown in \Fig{fig_phase}.

The method described above has been successfully applied to pressure modes in
\cite{Kastaun06,thesis}.
For inertial modes, the scheme is also usable, but computationally
less efficient.
The reason is that the ability of the Fourier analysis to separate
different frequencies and the corresponding eigenfunctions depends
on the time interval given by the evolution time.
For pressure modes, this is not a problem because their
frequencies are well separated.
For inertial modes, there seem to be infinitely
many modes inside a given frequency interval.
Although most of them are of high order and cannot be resolved by
simulations with a finite resolution,
the frequencies of resolvable modes can still become much closer
than those of the pressure modes, which makes it more difficult to
distinguish them by Fourier analysis.
Note that high order inertial modes can have similar frequencies as 
low order inertial modes, in contrast to pressure modes.

The direct approach is to use very long evolution times.
Since the usable evolution time is effectively limited by the timescale of the
numerical damping, one would also need higher resolutions.
The brute force method is not only computationally inefficient,
it is also not guaranteed to work, since with increasing resolution
more modes with similar frequencies are becoming resolvable.

However, 
there is a simpler method to compute at least the lower order modes,
by using the numerical errors causing the damping of the oscillations
to our advantage.
Higher order modes experience a stronger numerical damping, 
and are therefore suppressed by a certain factor during every mode 
recycling iteration, 
no matter how small the difference in frequency to the low order 
mode we want to extract might be.
This method to separate modes works best at medium resolutions 
around 150--200 points per stellar radius;
for higher resolutions,
the numerical damping becomes too weak, 
while for low resolutions,
the accuracy of the results is not sufficient.

In practice, we use up to ten mode recycling steps, evolution
times between 20--300$\usk\milli\second$, and resolutions between
150--300 points per stellar radius for inertial modes.
For pressure modes, one to three iterations, evolution times
around 10--20$\usk\milli\second$, and resolutions around 100 points
were usually sufficient.

The aforementioned complications also make it hard to properly compute the error
of the inertial mode eigenfunctions.
The error now depends not only on the accuracy of the simulation, 
which is investigated in \cite{Kastaun06,thesis},
but also on the extent to which modes with frequencies in the same
Fourier bin could be suppressed.
This cannot be estimated from the Fourier spectra.
However, if the eigenfunction of a neighboring mode present in the simulation
significantly differs from the mode of interest,
it would show up as an error in the uniformity of the global complex phase.
We cannot separate modes with very similar frequencies {\em and} eigenfunctions,
however.

Based on comparisons of different simulations of the same mode,
we estimate the error of the eigenfunctions to 5~\%.
Our estimate for the error of the frequency is
$\delta f = 0.01 f +  \Delta F$, where $\Delta F$ is the
width of the Fourier bin corresponding to the evolution time.

\subsection{Inertial mode eigenfunctions}
Using the method described in the previous section, we computed the frequencies and eigenfunctions
of five low order axisymmetric inertial modes 
for a sequence of rigidly rotating stars with various rotation rates and 
a fixed central density of $\Flt{7.9056}{17} \usk\kilo\gram\per\meter\cubed$.
The equation of state is a polytrope
\begin{equation}
  \frac{P}{\rho_p} = \left(\frac{\Rmd}{\rho_p}\right)^\Gamma,
\end{equation}
with $\Gamma=2$, $\rho_p = \Flt{6.1760}{18} \usk\kilo\gram\per\meter\cubed$.
Our stellar models are named BU2--BU4 and BU6 in \cite{SAF2004,DSF2006},
where the corresponding pressure modes are investigated.
Additionally, we use a model ``BUS'' with a slow rotation rate of $246 \usk\hertz$.
The fastest rotating model has a rotation rate of $792\usk\hertz$,
and an axes ratio of $0.7$.
For comparison: the mass-shedding limit for the given EOS and central
density is reached for an axis ratio of $0.58$.
Deviations from the slow rotation limit should therefore become visible.
We restricted ourselves to equatorially symmetric simulations,
in order to save computational resources.

\begin{figure}
  \includegraphics[]{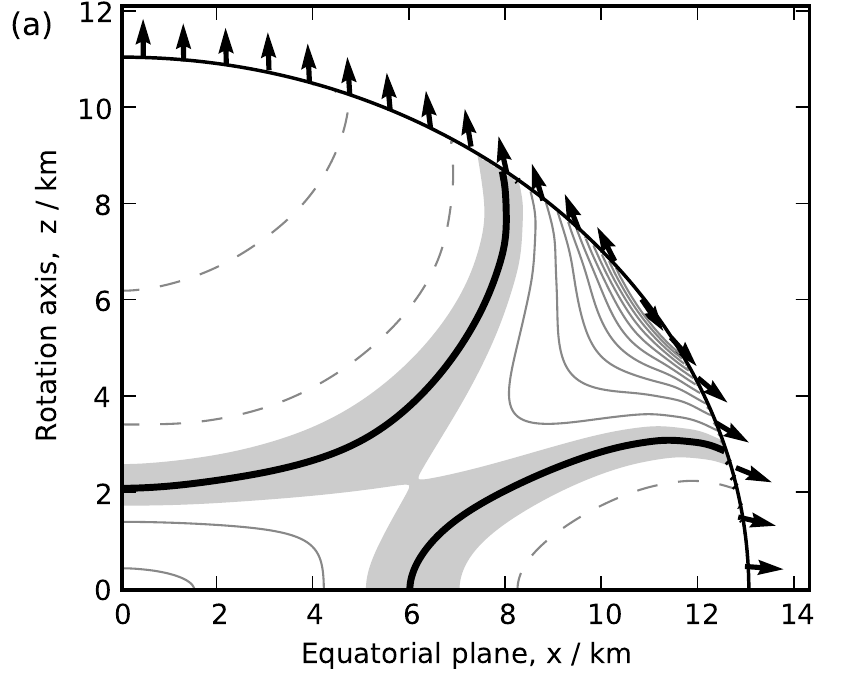}\\
  \includegraphics[]{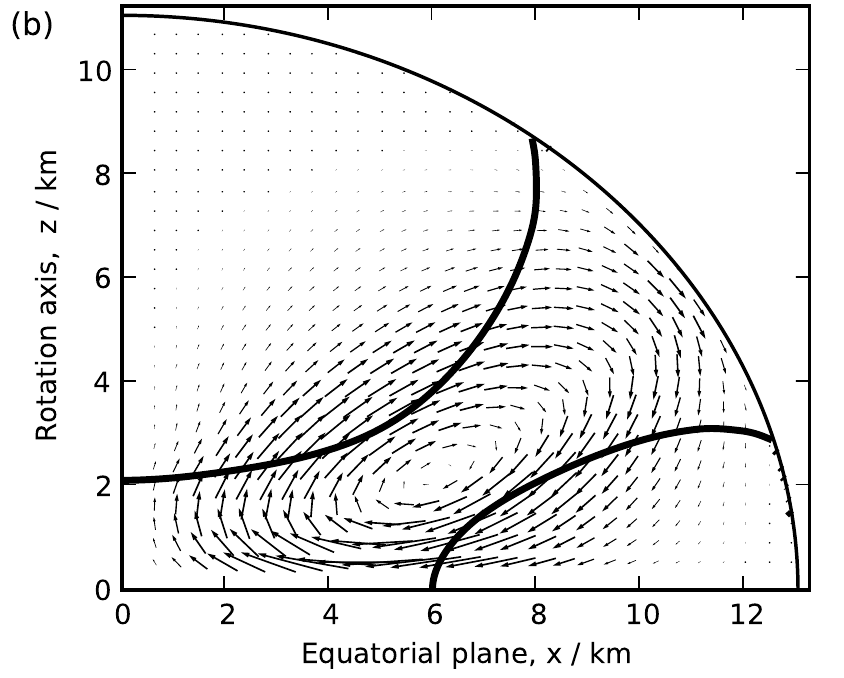}
  \caption{\label{fig_mode_i21}
    Eigenfunction of inertial mode $i_{21}$,
    computed for stellar model BU3.
    The cuts show the upper half of the meridional plane.
    (a) 
    Eigenfunction $\Ps$.
    With the exception of the star surface, the lines represent equidistant isocontours.
    Negative values correspond to dashed lines, and the thick solid line
    marks the node of the eigenfunction.
    The shaded area represents amplitudes below 5\% of the maximum one,
    which is our guess for the accuracy of the node location.
    The arrows at the surface represent the analytic prediction for the
    direction of the isopotential lines of $\Ps$ in the slow rotation approximation.
    (b)
    Eigenfunction of momentum density components $\Cmom^l$ in the meridional plane.
    The solid lines are the nodes of $\Ps$.
  }
\end{figure}

\begin{figure}
  \includegraphics[]{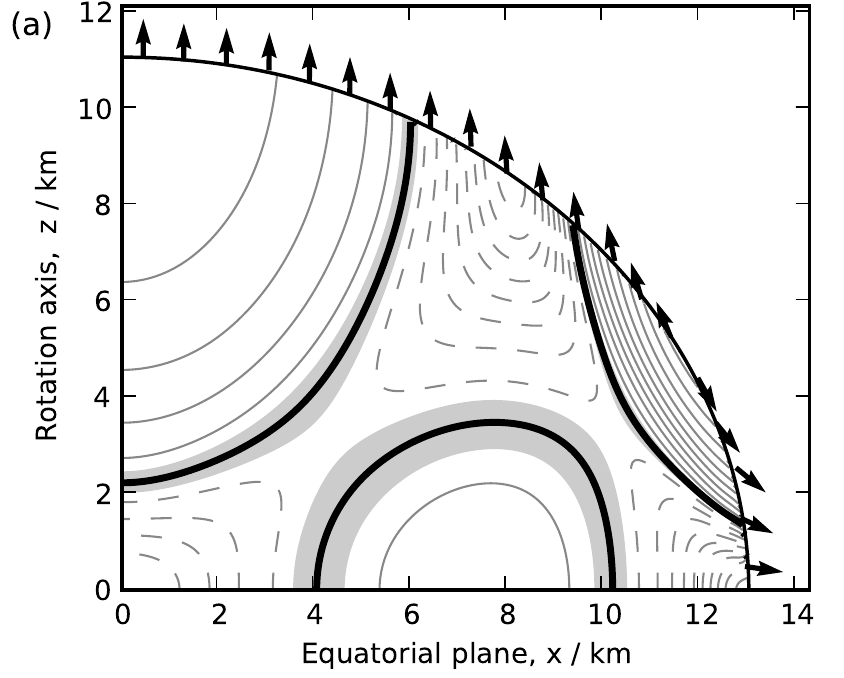}\\
  \includegraphics[]{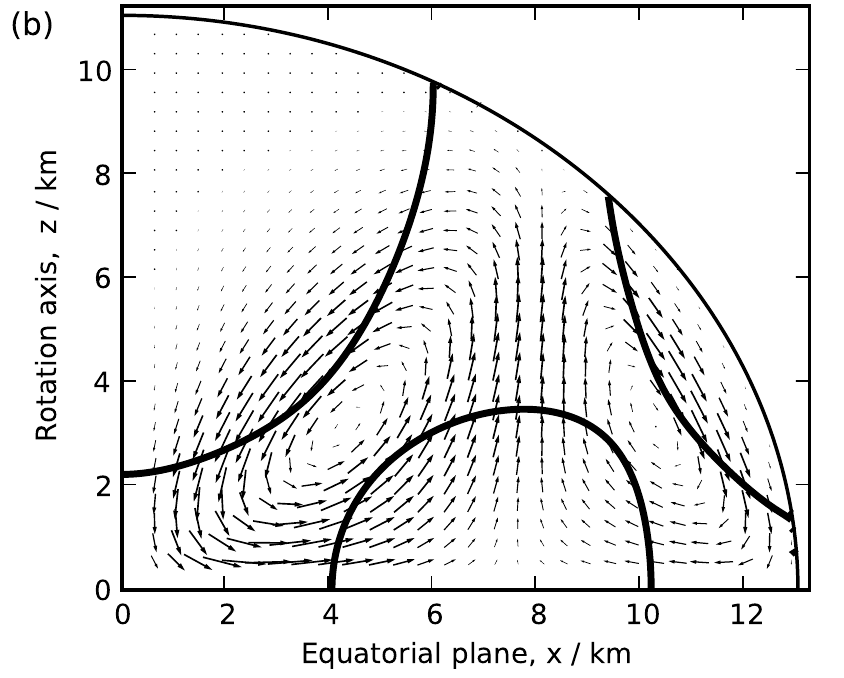}
  \caption{\label{fig_mode_i22}
    Like \Fig{fig_mode_i21}, but for the $i_{22}$ mode.
  }
\end{figure}

\begin{figure}
  \includegraphics[]{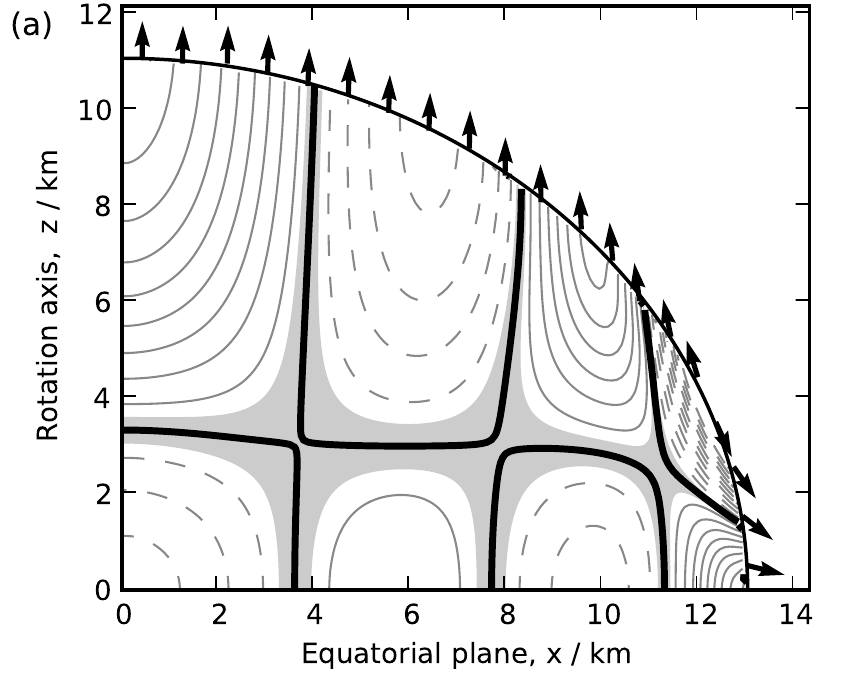}\\
  \includegraphics[]{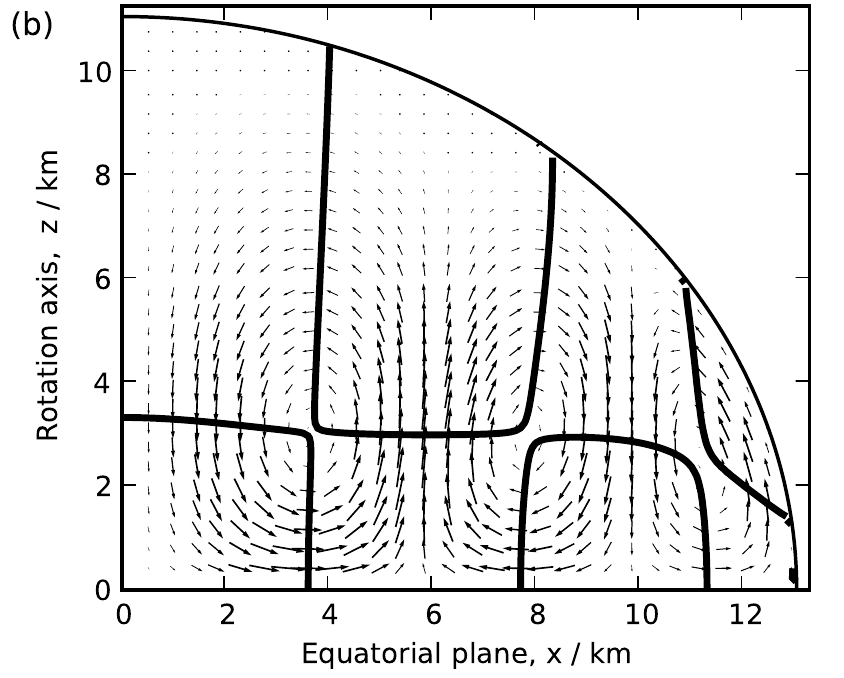}
  \caption{\label{fig_mode_i23}
    Like \Fig{fig_mode_i21}, but for the $i_{23}$ mode.
  }
\end{figure}

\begin{figure}
  \includegraphics[]{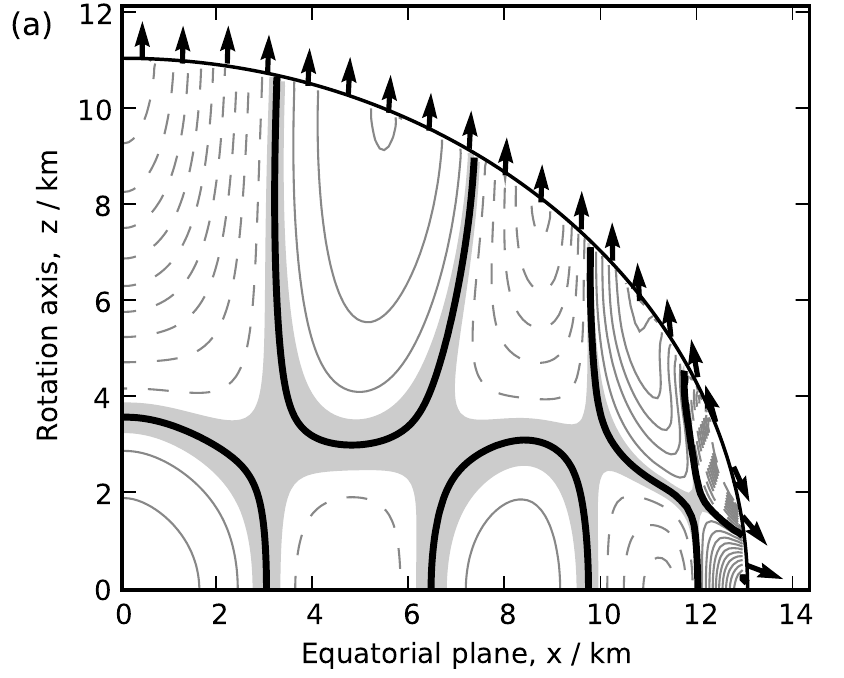}\\
  \includegraphics[]{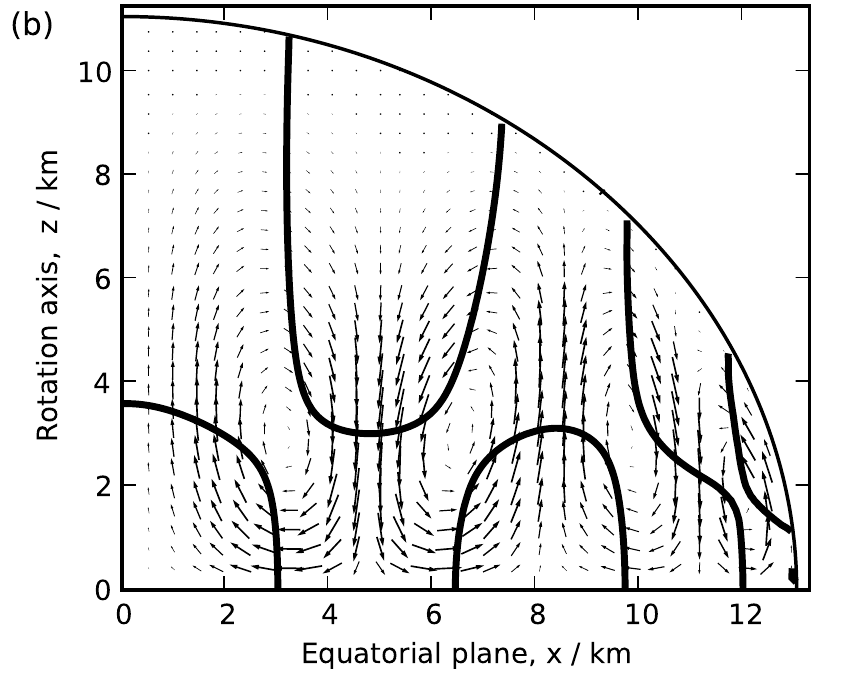}
  \caption{\label{fig_mode_i24}
    Like \Fig{fig_mode_i21}, but for the $i_{24}$ mode.
  }
\end{figure}

\begin{figure}
  \includegraphics[]{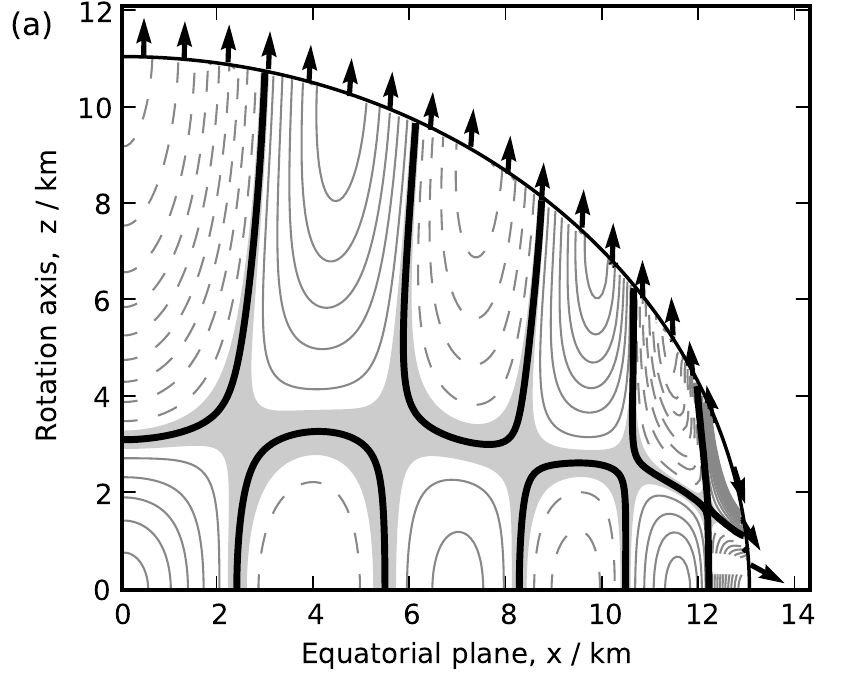}\\
  \includegraphics[]{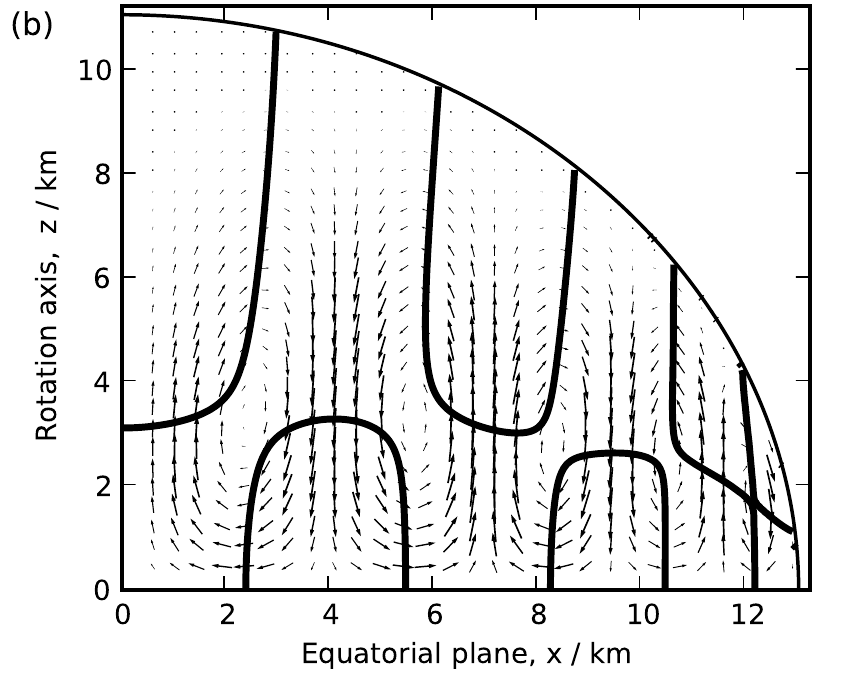}
  \caption{\label{fig_mode_i25}
    Like \Fig{fig_mode_i21}, but for the $i_{25}$ mode.
  }
\end{figure}

\begin{figure}
  \includegraphics[]{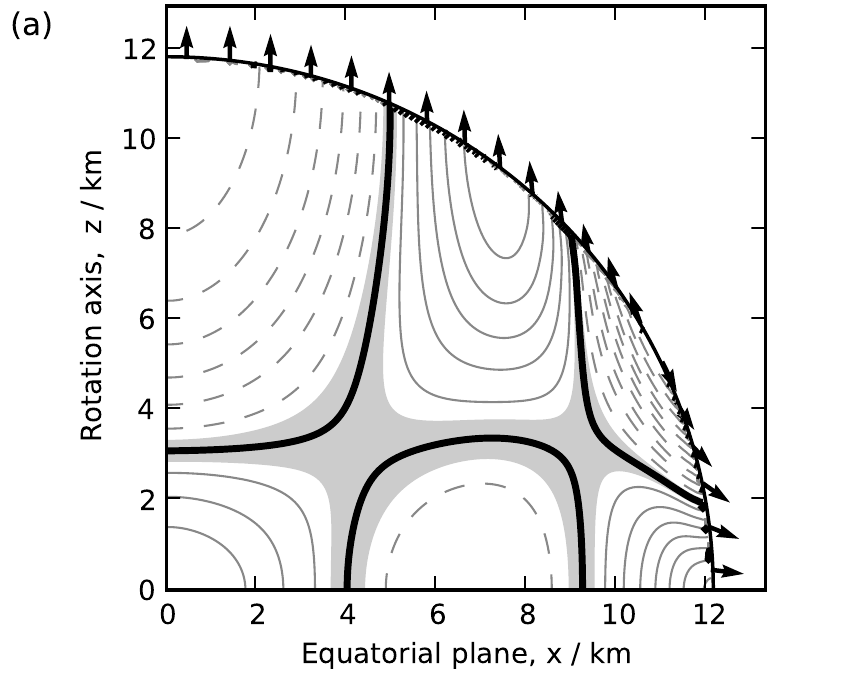}\\
  \includegraphics[]{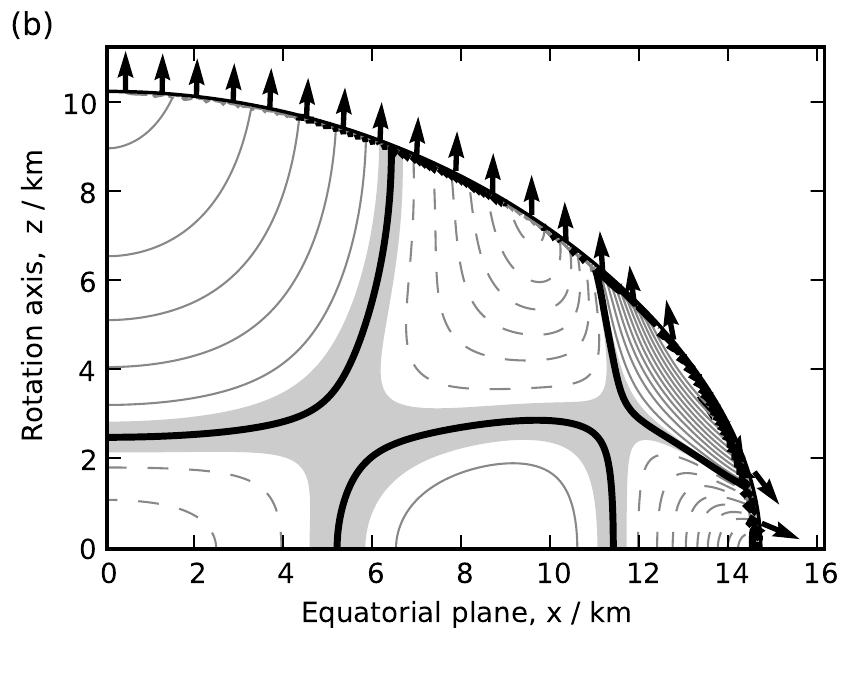}
  \caption{\label{fig_comp_i22}
    Comparison of the eigenfunction of inertial mode $i_{22}$
    at rotation rates $246\usk\hertz$ (a) and $792\usk\hertz$ (b).
    See \Fig{fig_mode_i21} for a description of the plot.
  }
\end{figure}

Our results for the axisymmetric eigenfunctions of model BU3,
which has a rotation rate of $590\usk\hertz$ and an axis ratio of $0.85$,
are shown in \Fig{fig_mode_i21} to \Fig{fig_mode_i25}.
As predicted in \Sec{sec_eeq}, inertial modes have a completely different structure
than pressure modes, at least for the axisymmetric case.
For comparison, pressure mode eigenfunctions for model BU3 can be found in \cite{thesis}.
The dependency of the patterns on the rotation rate is weak for all modes.
\Fig{fig_comp_i22} shows one of them at two different rotation rates.
Obviously, the angular dependency is  not given by a spherical harmonic plus
rotational corrections proportional to $\Omega^2$, like it is the case for
pressure modes.

For practical purposes, we classify the axisymmetric inertial modes by the number
of nodes of the eigenfunction $\Ps$ along the entire rotation axis and the equatorial
plane, e.g.\ $i_{23}$ has two nodes along the (entire) axis,
and three along the equatorial plane.
It is unclear wether the number of nodes in each direction uniquely determines
the mode; for the modes we extracted so far, this is the case.

For some simulations, the global complex phase remained a little fuzzy
in the regions where two nodes come close, as shown in \Fig{fig_phase}.
This could easily be explained by the fact that the phase is much more sensitive 
to errors in regions where the eigenfunction is nearly zero.
However we cannot rule out the possibility that our modes are
really superpositions of several modes with very similar frequencies 
whose eigenfunctions differ significantly only in the aforementioned regions.

Although we performed the mode recycling steps required to obtain a clean 
eigenfunction only for modes with two nodes along the axis,
due to limited computational resources, 
we also encountered patterns with more nodes along the axis.

The nodes of the extrated inertial mode eigenfunctions tend to be oriented
roughly parallel and orthogonal to the rotation axis,
which becomes more pronounced with increasing number of nodes
crossing the equatorial plane.
This corresponds well to the analytical prediction from \Sec{sec_wave}.

In order to check prediction \Eref{eq_ndir} for the direction of the gradient of $\Ps$
in the slow rotation limit, we computed $\Ndir^l$ using central finite differences
from the same background model data used as initial data in the simulations.
The result is compared to the numerical eigenfunctions in
\Fig{fig_mode_i21} to \Fig{fig_mode_i25}.
As one can see, the analytic prediction nicely matches the numerical results.

Another visible difference of the extracted inertial modes to pressure modes
is that the velocity fields are
vortexlike and tangential to the surface.
This reflects the fact that in the slow rotation approximation,
the density change vanishes, as shown in \Sec{sec_slow},
and the mass current must become divergence-free.

As a consequence, the oscillation amplitude $\Px^i$ of the fluid 
displacement vector can become arbitrary large in the slow rotation limit
without leaving the linear regime.
The fluid then performs convection-like motions which reverse periodically,
without changing the star profile.
Note that for a fixed velocity perturbation amplitude, the fluid
displacement vector $\Px^i$ \textit{has} to go to infinity in the slow rotation limit
due to \Eref{eq_ordering}.

\subsection{Inertial mode frequencies}
Together with the eigenfunctions, we extracted the frequencies.
In most cases, other modes were suppressed strongly enough to allow
a direct fit of an exponentially damped oscillation to the time evolution of
the velocity components at some sample point.
Otherwise, the frequencies were extracted from the corresponding
Fourier spectrum.

\begin{figure}
  \includegraphics[]{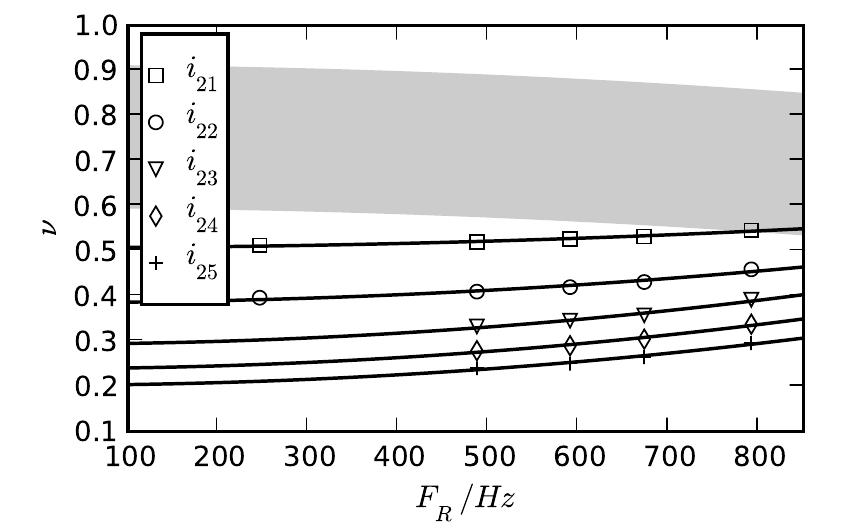}
  \caption{\label{fig_epsilons}
    Ratio $\En = \omega / 2 \Omega$ versus rotation rate $F_R$
    for a sequence of stellar models with fixed central density.
    The shaded region marks the range of $\Enc$
    inside the star. The solid lines are fits of the
    function $\En_0 + \En_2 \Omega^2$ to the data.
    The coefficients are given in \Tab{tab_fitfreq}.
  }
\end{figure}

The frequencies are given in \Tab{tab_freqs}.
As shown in \Sec{sec_eeq}, the frequencies of inertial modes should be
proportional to the rotation rate at leading order.
The actual ratio between the two depicted in \Fig{fig_epsilons} is indeed
compatible with a correction term to $\En$ of second order in the rotation rate.
By fitting the data accordingly, we obtain extrapolated values for the
slow rotation limit, which is inaccessible to our numerical method.
The resulting values are given in \Tab{tab_fitfreq}.
Note that our sequence of stellar models does not correspond to the same
star at different rotation rates since we fixed the central density
instead of the total baryonic mass.
However, the leading correction term to the total mass is also of
second order.

\begin{figure}
  \includegraphics[]{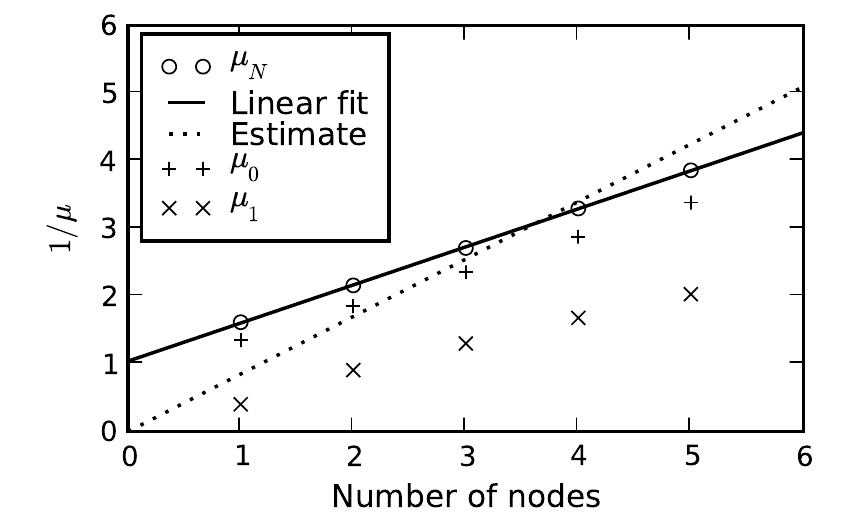}
  \caption{\label{fig_eps_order}
    Dependency of $1/\mu$ on the number of nodes crossing the
    equatorial plane, for the inertial modes of stellar model BU3 with
    exactly two nodes crossing the (entire) rotation axis.
    Plotted is the value in the Newtonian limit, where $\mu$ is
    computed setting $\Enc = 1$, as well as the minimum and maximum
    values $\mu_0$, $\mu_1$ in the actual stellar model.
    The solid line is a linear fit to the Newtonian values,
    the dotted line is the estimate \Eref{eq_approx} computed
    with $\Enc=1$.
  }
\end{figure}

\begin{table}
  \begin{ruledtabular}\begin{tabular}{c|c|c|.|.|.}
    Star model & $F_R / \hertz $  & Mode & 
    \multicolumn{1}{c|}{$f / \hertz$} & 
    \multicolumn{1}{c|}{$\delta f/f$} & 
    \multicolumn{1}{c}{$\En$} \\\hline
    BUS  & 246   & $i_{21}$ & 252.3 & 2.0 \usk\%& 0.5127 \\
         &       & $i_{22}$ & 195.2 & 2.8 \usk\%& 0.3967 \\ \hline
    BU2  & 487   & $i_{21}$ & 507.7 & 6.0 \usk\%& 0.5209 \\
         &       & $i_{22}$ & 399.8 & 1.9 \usk\%& 0.4102 \\
         &       & $i_{23}$ & 324.4 & 4.9 \usk\%& 0.3329 \\
         &       & $i_{24}$ & 272.1 & 4.7 \usk\%& 0.2792 \\
         &       & $i_{25}$ & 234.2 & 4.6 \usk\%& 0.2403 \\ \hline
    BU3  & 590   & $i_{21}$ & 621.8 & 1.6 \usk\%& 0.5262 \\
         &       & $i_{22}$ & 496.4 & 2.7 \usk\%& 0.4201 \\
         &       & $i_{23}$ & 408.6 & 7.2 \usk\%& 0.3458 \\
         &       & $i_{24}$ & 342.9 & 4.7 \usk\%& 0.2902 \\
         &       & $i_{25}$ & 296.2 & 6.7 \usk\%& 0.2507 \\ \hline
    BU4  & 673   & $i_{21}$ & 717.0 & 2.6 \usk\%& 0.5325 \\
         &       & $i_{22}$ & 581.0 & 3.9 \usk\%& 0.4315 \\
         &       & $i_{23}$ & 481.5 & 6.2 \usk\%& 0.3576 \\
         &       & $i_{24}$ & 410.4 & 7.1 \usk\%& 0.3048 \\
         &       & $i_{25}$ & 356.6 & 5.7 \usk\%& 0.2649 \\ \hline
    BU6  & 792   & $i_{21}$ & 865.3 & 6.8 \usk\%& 0.5462 \\
         &       & $i_{22}$ & 728.2 & 5.6 \usk\%& 0.4597 \\
         &       & $i_{23}$ & 621.2 & 3.1 \usk\%& 0.3921 \\
         &       & $i_{24}$ & 535.5 & 5.7 \usk\%& 0.3380 \\
         &       & $i_{25}$ & 469.0 & 4.6 \usk\%& 0.2960 \\
  \end{tabular}\end{ruledtabular}
  \caption{\label{tab_freqs}
    Frequencies $f$ of various inertial modes for the sequence
    of stellar models described in the main text.
    $\delta f$ is our estimate for the error of $f$.
    $F_R$ denotes the rotation rate of the star.
  }
\end{table}
The modes considered here satisfy an interesting empirical relation:
for a given stellar model,
there is a nearly linear relation between the quantity $1/\mu$ introduced in
\Sec{sec_eeq}, and the number of nodes along the equatorial plane.
Note that $\mu$ is not constant inside the star, and the accuracy
of the linear parametrization slightly depends on the choice of the
position where $\mu$ is computed (which has of course to be the same
for all modes).
For model BU3,
\Fig{fig_eps_order} shows the minimum and maximum values of $\mu^{-1}$
as well as the Newtonian value we define by setting $\Enc=1$.
The latter can be conveniently computed from the
rotation rate and frequency alone, without knowledge of the stellar model.
The relation also holds for the other stellar models and for the 
extrapolated values in the slowly rotating case,
but with a slightly different slope and offsets.

The analytical estimate \Eref{eq_approx} for higher order modes,
see \Sec{sec_wave},
which is also shown in \Fig{fig_eps_order}, matches the data only by order
of magnitude.
This is not surprising considering the drastic approximations
used in the derivation.
However, the fact that the data fulfills another linear relation
quite accurately
could be a hint for the existence of a better analytical estimate.

\begin{table}
  \begin{ruledtabular}
  \begin{tabular}{l|.|.|.|.|.}
     Mode & 
      \multicolumn{1}{c|}{$ i_{21} $} & 
      \multicolumn{1}{c|}{$ i_{22} $} & 
      \multicolumn{1}{c|}{$ i_{23} $} & 
      \multicolumn{1}{c|}{$ i_{24} $} & 
      \multicolumn{1}{c}{$ i_{25} $} \\\hline
     $\En_0$ &
       0.508 & 0.386 & 0.294 & 0.239 & 0.203 \\
     $\En_2 / \kilo\hertz^{-2}$ &
       0.058 & 0.109& 0.152& 0.153 & 0.145 \\
  \end{tabular}
  \end{ruledtabular}
  \caption{\label{tab_fitfreq}
    Coefficients of the parametrization $\En=\En_0 + \En_2 \Omega^2$,
    obtained by a fit to the numerical results for each inertial mode.
  }
\end{table}

As a further check, the $i_{21}$ inertial mode has been extracted by
\cite{DimPC08},
using the \Coconut code described in \cite{DimMdM,DSF2006}.
Since only a single mode recycling iteration was used, 
we estimate the errors of the eigenfunctions around 20\ \%,
judged from the observed phase errors and our previous mode recycling
experiences with inertial modes.
However, the eigenfunctions expose the same qualitative structure shown
in \Fig{fig_mode_i21}, allowing us to identify the mode.
The frequency of $634\usk\hertz$ matches our result within 2\ \%.

\clearpage

\section{Summary}
In this work, we extracted the frequencies and eigenfunctions
of several axisymmetric inertial modes of rigidly rotating
neutron star models with a polytropic equation of state,
for slow to medium rotation rates.
For this, we used a nonlinear code for the evolution of ideal
fluids in arbitrary spacetimes in conjunction with the mode
recycling technique for the extraction of single oscillation modes.
The spacetime is kept fixed, i.e.\ we use the Cowling approximation.

The extracted frequencies are proportional to the rotation rate at leading
order, and well below two times the rotation rate.
The spectrum of mode frequencies seems quite dense, which made
it more difficult to extract single inertial modes, compared to pressure modes.
Although the slow rotation limit is not accessible to our numerical method,
we computed the frequencies in the slow rotation limit using extrapolation.

The scalar eigenfunctions of the axisymmetric inertial modes exhibit a checkerboard like
structure.
In contrast to pressure modes, it seems unnatural to classify axisymmetric
inertial modes by the dominant term of a decomposition into spherical harmonics;
which term is dominant depends on the radius, 
and also globally there is no strongly dominant term.
That does not mean such decomposition is not a useful technique,
but one should keep it in mind when talking e.g.\ of a $l=2, m=0$ inertial mode.
Note the former might not apply for nonaxisymmetric modes,
which we did not extract.
For practical purposes, we used the number of nodes
along equatorial plane and rotation axis to classify the axisymmetric 
modes.

In the analytic part of our work, we derived a simple scalar eigenequation
describing the eigenfunctions in the case of rigid rotation.
Using this equation, we investigate the behavior of inertial modes at the
surface, which agrees well with our numerical results.
Further, we derived an approximate relation between the frequency of the mode
and the number of nodes along the equatorial plane and the rotation axis.
This relation predicts that the frequency grows with the number
of nodes along the rotation axis, but decreases with the number of nodes
along the equatorial plane.
The latter indeed holds for the numerical results.
The first part has not been validated yet since so far we
systematically investigated only modes with two nodes along the axis.

The relation was derived for the case of higher order modes,
and even then the expected accuracy is around 30\ \% at most.
Surprisingly, the numerical results fulfill a similar relation
quite accurately, even for the lowest order modes.
This leads us to speculate wether this empirical relation could
possibly be derived analytically.

We also investigated the slow rotation limit.
Since inertial mode frequencies are proportional to the rotation rate,
in contrast to pressure modes,
we obtain two different eigenequations describing pressure and inertial modes.
The difference is due to terms related to the Coriolis force,
which remains dominant in the slow rotation limit for inertial modes,
but becomes negligible for pressure modes.
Accordingly, the second order scalar partial differential eigenequation 
we obtain in that limit is invariant under any rotation for pressure modes,
but for inertial modes there is a preferred direction given by the rotation
axis.

We have shown that in the slow rotation limit, 
gravitational radiation of inertial modes in isentropic stars
is mainly due to the mass currents instead
of the density perturbations, 
at least on the level of the quadrupole formula. 
Further, the gravitational radiation should become negligible for 
the dynamics in the slow rotation limit, as it's luminosity
goes like $\Omega^6$ for a fixed kinetic energy of the oscillation.

We have no quantitative estimate how strongly inertial modes are affected 
by neglecting the perturbation of the gravitational potential.
The available results e.g.\ of \cite{DSF2006} for pressure modes cannot be 
applied to inertial modes since the relations between the magnitudes of 
different quantities are completely different.
To estimate the accuracy of the Cowling approximation for inertial modes,
we need fully relativistic studies.

Since the axisymmetric inertial modes investigated here 
are not excited by rotational instabilities and are weak emitters
for all but the most rapidly rotating stars, 
for which their frequency becomes comparable to those of pressure modes,
a direct detection of the resulting gravitational waves seems unlikely.
We are also not aware of any other mechanism which would allow to detect
those modes. However, given our limited knowledge of neutron star physics,
this might quickly change.
We believe it is important to know the complete mode spectrum
of neutron stars for future studies,
e.g.\ of mode coupling effects
or the interaction of oscillations with the crust and the magnetosphere.

Finally, we gave heuristic arguments why inertial modes should have a different
structure in the presence of an entropy gradient, but only if the pressure
perturbation of the isentropic oscillation
becomes small in comparison to the one caused by the entropy gradient; 
as a consequence, oscillations with a given kinetic energy are more sensible to
entropy gradients for slowly rotating stars.

\begin{acknowledgments}
This work was supported by the collaborative research
center `SFB-Transregio 7
Gra\-vi\-ta\-tions\-wellen\-astronomie''
of the German Science Foundation (DFG).
Computations have been carried out mainly on the ``Pioneer''
cluster of the University of T\"ubingen.
Thanks go to Harald Dimmelmeier and Kostas Kokkotas for
fruitful discussions.
\end{acknowledgments}

\bibliography{article.bib}

\end{document}